\documentclass[fleqn,twoside,twocolumn,prb]{revtex4}  

\usepackage{times,float}
\usepackage{amsmath}
\usepackage{psfrag,epsfig}

\def\picdirectory{.}

\def\fl{\hspace*{-\mathindent}}

\def\ep{\varepsilon}

\def\Name#1{#1,}
\def\Book#1{\textit{#1}}
\def\Review#1#2#3#4{#1 \textbf{#2}, #4 (#3)}
\def\Publ#1#2{(#1 #2)}
\def\Editor#1{ed. #1}

\begin{document}

\title{Rheology of gelling polymers in the Zimm model}

\author{Henning L\"owe}
\email{loewe@theorie.physik.uni-goettingen.de}

\author{Peter M\"uller}
\email{mueller@theorie.physik.uni-goettingen.de}

\author{Annette Zippelius}
\email{zippelius@theorie.physik.uni-goettingen.de}

\affiliation{Institut f\"ur Theoretische Physik,
         Georg-August-Universit\"at, D--37077 G\"ottingen, Germany}

\date{\today}

\begin{abstract}
  In order to study rheological properties of gelling systems in
  dilute solution, we investigate the viscosity and the normal
  stresses in the Zimm model for randomly crosslinked monomers. The
  distribution of cluster topologies and sizes is assumed to be given
  either by Erd\H os-R\'enyi random graphs or three-dimensional bond
  percolation. Within this model the critical behaviour of the
  viscosity and of the first normal stress coefficient is determined
  by the power-law scaling of their averages over clusters of a given
  size $n$ with $n$. We investigate these Mark--Houwink like scaling
  relations numerically and conclude that the scaling exponents are
  independent of the hydrodynamic interaction strength. The
  numerically determined exponents agree well with experimental data
  for branched polymers. However, we show that this traditional model
  of polymer physics is not able to yield a critical divergence at the
  gel point of the viscosity for a polydisperse dilute solution of
  gelation clusters. A generally accepted scaling relation for the
  Zimm exponent of the viscosity is thereby disproved.
\end{abstract}

\pacs{82.70.Gg  36.40.Sx  83.60.Bc}

\maketitle

%
%
\section{Introduction}\label{Sec1}
%
%

The influence of hydrodynamic interactions on critical rheological properties
of gelling polymeric systems has been discussed controversely for many
decades. In particular, the experimental values for the exponent $k$, which
governs the divergence of the shear viscosity $\eta\propto \ep^{-k}$ with
vanishing distance $\ep$ to the critical gel point, scatter considerably, see
Table~\ref{tab:1} for some examples. In order to interpret the wide scatter of
the data, they are often related to either Zimm or Rouse dynamics, depending
on whether hydrodynamic interactions are believed to be relevant or not. In
this paper we intend to elaborate on the validity of this interpretation, so
let us be precise with the labels. By definition, the {\it Rouse}
model\cite{Rou53} neglects both hydrodynamic and excluded-volume interactions.
Its straightforward generalization from linear polymers to a gelling melt of
randomly crosslinked monomers provides a microscopic framework, within which
one can derive an exact scaling relation for the viscosity exponent.
\cite{BrLoMu99,BrLoMu01a, BrLoMu01b,BrMuZi02, Mul03} The {\it Zimm} model,
\cite{Zim56} by definition, takes hydrodynamic interactions into account on a
preaveraged level, but still neglects excluded-volume interactions. We are not
aware of a microscopic approach based on the Zimm model which allows for an
exact analytic computation of the viscosity exponent for a gelling polymeric
solution.  Other models for gelling polymers, which go beyond the Rouse or
Zimm model by incorporating excluded-volume effects or fluctuating
hydrodynamic interactions, are conjectured to belong to {\it different}
universality classes and will not be considered here.

\begin{table}
  \centering
  \begin{tabular}{c||c|c|c|c|c|c|c|c}
    $k$   & ~$0.2$~ & ~$0.79$~ &  ~$0.82$~ & ~$0.95$~ & ~$1.27$~ & ~$1.36$~ &
    ~$>1.4$~ & ~$6.1$~ \\ \hline
    Ref.~ & \onlinecite{TaUrMa90} & \onlinecite{AdDeOk79} & 
    \onlinecite{AxKo90} & \onlinecite{DuBa80} & \onlinecite{ZhZhJi96} &
    \onlinecite{LuMoWi95} & \onlinecite{CoGiRu93} & \onlinecite{LuMoWi99}
  \end{tabular}
  \caption{Experimental data for the critical exponent $k$ of the viscosity 
    at the gelation transition.}
  \label{tab:1}
\end{table}

Scaling theory \cite{StCoAd82,Mut85,Cat85} has proven to be a powerful tool to
describe the properties of polymeric systems. The relaxation time $t_n$ of a
typical cluster of $n$ monomers (henceforth $n$ will be referred to as the
\emph{size} of the cluster) is estimated to be $t_n\sim R_n^2/D_n$, where
$R_n$ is the radius of gyration and $D_n$ the diffusion constant of the
cluster. The scaling $R_n\sim n^{1/d_f}$ of the radius of gyration of a
cluster with size $n$ is determined by the Hausdorff fractal dimension $d_f$.
The diffusion constant is assumed to scale like $D_n\sim 1/R_n$ in the Zimm
model.  This assumption is based on the Stokes Einstein relation which is
valid for linear polymers and still holds for the diffusion of fractal polymer
clusters in the Zimm model. \cite{KuLoMu03} The average contribution of 
clusters of size $n$ to the viscosity is then given by $\eta_n\sim t_n/n$.
This implies the scaling \cite{StCoAd82,Mut85,Cat85}
\begin{equation}
  \label{zimmscaling}
  \eta_n \sim n^{b_{\eta}},\quad\quad  b_{\eta}=d/d_{f}-1\
\end{equation}
where $d$ is the spatial dimension. With an underlying distribution of
cluster sizes, which is widely believed to follow the scaling laws of
percolation, this gives rise to the exponent $k=(1-\tau+d/d_f)/\sigma$
for the averaged viscosity in terms of the static percolation
exponents.

Beside critical properties, recent publications aim at the dynamics of
single clusters with particular topologies within the Zimm model.
Refs.~\onlinecite{ZiKi96} and~\onlinecite{FeBl02} consider the Zimm
dynamics of star-shaped clusters and dendrimers, and
Ref.~\onlinecite{JuKoBl03} analyses the relaxation behaviour of
fractal (Sierpinski-type) clusters in the Zimm model. The latter
authors mention the possibility of non-universal behaviour. The
question of non-universality is also raised from computer simulations
\cite{PlVeJo03} of gelling liquids under the influence of solvent
particles.

In this paper we investigate the viscosity and the normal stresses in the Zimm
model for randomly crosslinked monomers. The distribution of cluster
topologies and sizes is assumed to be given either by Erd\H{o}s--R\'enyi
random graphs or three-dimensional bond percolation.  The details of the model
are described in Section~\ref{Sec2}. Within this model, the critical behaviour
of the viscosity $\eta$ and of the first normal stress coefficient
$\Psi^{(1)}$ is determined by the scaling with $n$ of the partial averages
${\eta}_n\sim n^{b_{\eta}}$, respectively $\Psi^{(1)}_n\sim n^{b_{\Psi}}$,
over clusters of size $n$ (Section~\ref{Sec3}). In Section~\ref{Sec4} we
investigate these Mark--Houwink like scaling relations numerically for
different strengths of the hydrodynamic coupling constant. We conclude in
Section~\ref{Sec5} that ~(i)~these scaling relations are governed by universal
exponents ${b_{\eta}}$ and ${b_{\Psi}}$. This conclusion is substantiated by
comparing our results to those for ring polymers in the Zimm model, which are
known to exhibit universal behaviour.  (ii)~~We find that the scaling relation
(\ref{zimmscaling}) does not agree with our numerical data and, hence, does
not describe the viscosity in the Zimm model for randomly crosslinked
monomers.


%
%
\section{Dynamic Model and its solution}\label{Sec2}
%
%

\subsection{Hydrodynamic Interactions}\label{Sec2.1}
We consider $N$ point-like monomers, which are characterized by their
time-dependent position vectors $\mathbf{R}_i(t)$, $i=1,\ldots, N$, in
three-dimensional Euclidean space. Permanently formed
crosslinks constrain $M$ randomly chosen pairs of particles 
$(i_e, j_e)$, $e=1,\ldots,M$. We study the dynamics of crosslinked
monomers in the presence of a solvent fluid, giving rise to
hydrodynamic interactions between the monomers. Purely relaxational
dynamics in an incompressible fluid subjected to an external space-
and time-dependent flow $\mathbf{v}(\mathbf{r},t)$ is
described by the equation of motion\cite{BiCu87,DoEd88} 
\begin{multline}
  \label{dynamics}
  \fl
  \frac{{d}}{{d}t}\mathbf{R}_i(t) - \mathbf{v}(\mathbf{R}_i(t),t) \\ 
  \fl = 
  \sum_{j=1}^{N} \boldsymbol{\mathsf{H}}_{i,j} \bigl(\mathbf{R}_{i}(t) -
  \mathbf{R}_{j}(t)\bigr) \Bigl( -\frac{\partial V}{\partial
    \mathbf{R}_j(t)} \Bigr)  + \mathbf{f}_i(\mathbf{R}_i(t),t)\,.
\end{multline}
Here, crosslinks are modelled by Hookean springs in the potential energy
\begin{equation} 
  \label{energy} 
  V := \frac{3}{2a^2}\:\sum_{e=1}^M 
  \bigl( \mathbf{R}_{i_e}-\mathbf{R}_{j_e} \bigr)^2 
  =: \frac{3}{2a^2}\: \sum_{i,j=1}^{N}\mathbf{R}_{i}\cdot
  {\Gamma}_{i,j}\,\mathbf{R}_{j}\,,
\end{equation}
where the length $a>0$ plays the role of an inverse crosslink strength and
physical units have been chosen such that $k_{\mathrm{B}}T=1$.
A given crosslink configuration $\mathcal{G}=\{i_e,j_e\}_{e=1}^M$ is
specified by its $N\times N$-connectivity matrix $\Gamma$.
Moreover, we impose a simple shear flow 
\begin{equation}
  \label{shearflow}
  \mathbf{v}(\mathbf{r},t) :=   
  \left(
    \begin{array}{ccc}
      0&\dot{\gamma}(t) & 0\\
      0&0&0\\
      0&0&0\\
    \end{array}
  \right) \, \mathbf{r},
\end{equation}
which is characterized by its time-dependent shear rate $\dot{\gamma}(t)$.
The mobility matrix is given by
\begin{equation}
  \label{mobility}
  \fl
  \boldsymbol{\mathsf{H}}_{i,j}(\mathbf{r}) := \delta_{i,j}
  \;\frac{1}{\zeta} \;  \boldsymbol{\mathsf{1}} + 
  (1- \delta_{i,j}) \; \frac{1}{8\pi\eta_{s}|\mathbf{r}|}\; 
  \biggl( \boldsymbol{\mathsf{1}} + 
  \frac{\phantom{^{\cdot}}\mathbf{r}
    \mathbf{r}^{\dagger}}{|\mathbf{r}|^{2}} \biggr). 
\end{equation}
The diagonal term in (\ref{mobility}) accounts for a 
frictional force with friction constant $\zeta$ that acts when a
monomer moves relative to the externally imposed flow field
(\ref{shearflow}). The non-diagonal term reflects the influence of the
motion of monomer $j$ on the solvent at the position of monomer
$i$ and is given by the Oseen tensor.\cite{Ose10,KiRi48} Here
$\eta_s$ denotes the solvent viscosity, $\delta_{i,j}$ the Kronecker
symbol, $\boldsymbol{\mathsf{1}}$ the 
three-dimensional unit matrix and the dagger indicates the
transposition of a vector. Rouse dynamics is recovered, if the
non-diagonal terms $i \neq j$ of the mobility matrix are neglected.
The Gaussian thermal-noise force fields $\mathbf{f}_{i}(\mathbf{r},t)$ in
(\ref{dynamics}) have zero mean and covariance 
\begin{equation}
  \label{covariance}
  \overline{\mathbf{f}_i(\mathbf{r},t)
    \,\mathbf{f}_{j}^{\dagger}(\mathbf{r}',t')} = 2\, 
  \boldsymbol{\mathsf{H}}_{i,j}(\mathbf{r}-\mathbf{r}')
  \,\delta(t-t')\,.
\end{equation}
Here $\delta$ stands for the Dirac-delta function and
the overbar indicates 
the Gaussian average over all realizations of $\mathbf{f}$.

In order to determine the model completely, it only remains to specify the
probability distribution of the crosslink configurations. We shall discuss
two different types of probability distributions: 
\hspace{1ex}(i)\hspace{1ex} crosslinks are chosen independently with
equal probability for every pair 
of monomers, corresponding to Erd\H{o}s--R\'enyi random graphs,
\cite{ErRe60} and \hspace{1ex}(ii)\hspace{1ex}
a distribution of crosslinks, which generates clusters amenable to the
scaling description of finite-dimensional percolation. \cite{StAh94}
The precise characterization of these distributions is given below.

\subsection{Preaveraging Approximation}\label{Sec2.2}

The equation of motion (\ref{dynamics}) is nonlinear due to the nonlinear
dependence of the mobility on the particles' positions. A simple but
uncontrolled approximation is the so-called preaveraging approximation
that was first introduced by Kirkwood and Riseman \cite{KiRi48} and
Zimm. \cite{Zim56} In this approximation the mobility matrix
(\ref{mobility}) is replaced by its expectation value
$\langle\boldsymbol{\mathsf{H}}_{i,j}\rangle_{\mathrm{eq}}$, which is
computed with respect to the equilibrium  
distribution, i.e.\ the Boltzmann weight $\sim e^{-V}$. Due to
rotational invariance of the potential (\ref{energy}),
the averaged mobility matrix is a multiple of the identity matrix
$\langle\boldsymbol{\mathsf{H}}_{i,j}(\mathbf{R}_{i}-\mathbf{R}_{j})
\rangle_{\mathrm{eq}}=\mathsf{H}^{\mathrm{eq}}_{i,j}\,\boldsymbol{\mathsf{1}}$,  
where  
\begin{equation}
  \label{preav1}
  \mathsf{H}^{\mathrm{eq}}_{i,j}:=
  \delta_{i,j}\;\frac{1}{\zeta}
  +(1-\delta_{i,j})\;\frac{1}{6\pi\eta_{s}}
  \left\langle\frac{1}{|\mathbf{R}_{i}-\mathbf{R}_{j}|}
  \right\rangle_{\mathrm{eq}}. 
\end{equation}
In the computation of (\ref{preav1}),
care has to be taken of the zero eigenvalues of the connectivity
matrix, corresponding to the translation of whole clusters.
To this end we regularize the potential (\ref{energy}) by adding a
confining term $3\omega/(2a^{2})
\sum_{i=1}^N\mathbf{R}_{i}\cdot\mathbf{R}_{i}$ and letting 
$\omega>0$ tend to zero subsequently. The average in (\ref{preav1}) is
conveniently performed via 
the Fourier representation of $1/|\mathbf{r}|$, and the result
\begin{align}
  \label{preav2}
  \left\langle\frac{1}{|\mathbf{R}_{i}-\mathbf{R}_{j}|}
  \right\rangle_{\mathrm{eq}}   
  = \frac{1}{a}\sqrt{\frac{6}{\pi}} \; \lim_{\omega\downarrow 0}
  \bigg( & [\mathsf{G}(\omega)]_{i,i}  + [\mathsf{G}(\omega)]_{j,j} \nonumber\\
  & -2[\mathsf{G}(\omega)]_{i,j}\bigg)^{-1/2}
\end{align}
involves the resolvent $\mathsf{G}(\omega):=({\Gamma}+\omega\mathsf{1})^{-1}$
of ${\Gamma}$.  The limit ${\omega\downarrow 0}$ is taken by expanding the
resolvent $\mathsf{G}(\omega)=\mathsf{E}_0/\omega+\mathsf{Z}+{\cal O}(\omega)$
in terms of $\omega$. Here $\mathsf{Z} := (\mathsf{1}-\mathsf{E}_0)/{\Gamma}$
is the Moore--Penrose inverse \cite{Al72} of the connectivity matrix, {\it
  i.e.}\ the inverse of ${\Gamma}$ restricted to the subspace of non-zero
eigenvalues. Moreover, $\mathsf{1}$ denotes the $N\times N$-unit matrix and
$\mathsf{E}_0$ the projector on the nullspace of ${\Gamma}$, which is spanned
by the vectors that are constant when restricted to any one cluster of
crosslinked monomers. More precisely, the matrix element
$[\mathsf{E}_0]_{i,j}$ is given by the inverse number of monomers of the
cluster if $i$ and $j$ are in the same cluster and zero otherwise (cf.\ Sec.\ 
II.D in Ref.~\onlinecite{BrLoMu01a} for details). Hence, the right-hand side
of (\ref{preav2}) vanishes for $\omega\downarrow 0$ whenever $i$ and $j$
belong to different clusters. Consequently, the preaveraged mobility matrix
$\mathsf{H}^{\mathrm{eq}}$ shows correlations of different particles only if
these particles are in the same cluster, in other words it is block-diagonal
and within one block given by
\begin{equation} 
  \label{preav}
  \mathsf{H}^{\mathrm{eq}}_{i,j}=
  \frac{1}{\zeta}\;\Bigl[ \delta_{i,j}
  +(1-\delta_{i,j})\,
  h\left(\kappa^2\,\pi/\mathcal{R}_{i,j} \right)\Bigr]\,.
\end{equation}
For convenience we introduced the function $h(x)=\sqrt{x/\pi}$ and the
quantity $\mathcal{R}_{i,j} :=
\mathsf{Z}_{i,i}+\mathsf{Z}_{j,j}-2\mathsf{Z}_{i,j}$, which can be interpreted
as the resistance between nodes $i$ and $j$ in a corresponding electrical
resistor network. \cite{KlRa93} The parameter $\kappa :=
\sqrt{6/\pi}\,\zeta/(6\pi\eta_{s}a)$ plays the role of the coupling constant
of the hydrodynamic interaction. Note that this definition of $\kappa$ differs
from that of other authors by a factor of $\sqrt{2}$, \cite{OeZy92}
respectively $\sqrt{6}/\pi$.  \cite{JuKoBl03} Formally setting $\kappa=0$ in
(\ref{preav}) yields $\mathsf{H}^{\mathrm{eq}}_{i,j} = \zeta^{-1}
\delta_{i,j}$, and the Zimm model for gelation reduces to the Rouse model for
gelation.\cite{BrGoZi97,BrLoMu99,BrLoMu01a, BrLoMu01b, BrMuZi02, Mul03} It is
well known that the Oseen tensor does not give rise to a positive-definite
mobility matrix for all possible spatial configurations of monomers. This
defect is cured if the Rotne--Prager--Yamakawa tensor\cite{RoPr69,Yam70} is
used instead.  Again, the preaveraging procedure is done with a confining
potential which is switched off afterwards. The function $h$ is then given
by\cite{Fix83}
\begin{equation}
  \label{preavrotne}
  h(x)=\mathrm{erf}(\sqrt{x})-\frac{1}{\sqrt{\pi}}\frac{1-\exp(-x)}{\sqrt{x}}.
\end{equation} 
It involves the error function $\mathrm{erf}(x)$ and recovers the
form of the preaveraged Oseen-Tensor asymptotically as $x \downarrow 0$.
As a result of preaveraging we obtain the \emph{Zimm model for
  crosslinked monomers in solution} 
\begin{equation} 
  \label{zimm2}
  \frac{{d}}{{d}t} \mathbf{R}_{i}(t) -
  \mathbf{v}(\mathbf{R}_{i}(t),t)   = 
  -\sum_{j=1}^{N} \mathsf{H}^{\mathrm{eq}}_{i,j}\:
  \frac{\partial V}{\partial \mathbf{R}_{j}(t)} 
  + \boldsymbol{\xi}_{i}(t) \,.
\end{equation}
Here, the covariance of the thermal noise is given by
\begin{equation}
  \overline{\boldsymbol{\xi}_{i}(t)
    \,\boldsymbol{\xi}_{j}^{\dagger}(t')} =
  2\,\mathsf{H}^{\mathrm{eq}}_{i,j} \,\delta(t-t') \boldsymbol{\mathsf{1}} \,.
\end{equation}
Since both the connectivity matrix $\Gamma$ and the preaveraged
mobility matrix $\mathsf{H}^{\mathrm{eq}}$ are block-diagonal, it follows
that clusters move {\it independently} of each other in this model.

\subsection{Formal Solution}\label{Sec2.3}

The Zimm equation (\ref{zimm2}) is linear, hence it can be solved
exactly. This is most conveniently done by introducing new coordinates
$\widetilde{\mathbf{R}}_{i}(t)$ through the coordinate transformation
\begin{equation}
  \label{koord}
  {\mathbf{R}}_{i}(t) =: \sum_{j=1}^N\left[\left(
      \mathsf{H}^{\mathrm{eq}}\right)^{1/2}\right]_{i,j}
  \widetilde{\mathbf{R}}_{j}(t) \,.    
\end{equation}
The resulting equation of motion for $\widetilde{\mathbf{R}}_{i}(t)$
coincides with that of the Rouse 
model for crosslinked monomers, \cite{BrGoZi97,BrLoMu99,BrLoMu01a, BrLoMu01b,
  BrMuZi02, Mul03} if 
one replaces the connectivity matrix $\Gamma$ by
\begin{equation}
  \label{gammatilde}
  \widetilde{{\Gamma}} := 
  (\mathsf{H}^{\mathrm{eq}})^{1/2} {\Gamma}\,
  (\mathsf{H}^{\mathrm{eq}})^{1/2}
\end{equation}
in the latter. Different coordinate transformations are commonly used
to establish this formal relation between the two models. We prefer
(\ref{koord}), because then the transformed equation of motion
involves the {\it 
  symmetric} matrix (\ref{gammatilde}). The resulting monomer
trajectories for (transformed) initial data
$\widetilde{\mathbf{R}}_{i}(t_0)$ are therefore given by 
\begin{align}
  \label{solution}
  \widetilde{\mathbf{R}}_{i}(t)
  =  \sum_{j=1}^N \biggl\{ & \widetilde{\mathsf{U}}_{i,j}(t-t_0)\,
  \boldsymbol{\mathsf{T}}(t,t_0)\, 
  \widetilde{\mathbf{R}}_{j}(t_0) \nonumber\\
  & +\int_{t_0}^{t} {d}t'\,\widetilde{\mathsf{U}}_{i,j}(t-t')\,
  \boldsymbol{\mathsf{T}}(t,t')\, \widetilde{\boldsymbol{\xi}}_{j}(t')\biggr\}\,,
\end{align}
as follows e.g.\ from Sec.~II.C in Ref.~\onlinecite{BrLoMu01a}.
The solution (\ref{solution}) is expressed in terms of the
transformed thermal noise with zero mean and covariance
\begin{equation}
  \overline{\widetilde{\boldsymbol{\xi}}_{i}(t)
    \,\widetilde{\boldsymbol{\xi}}_{j}^{\dagger}(t')} =
  2\,\delta_{i,j} \,\delta(t-t')\boldsymbol{\mathsf{1}} \,,
\end{equation}
and the time evolution in the simple shear flow (\ref{shearflow}) is
characterized by the $N\times N$-matrix 
\begin{equation}
  \widetilde{\mathsf{U}}(t):=\exp\bigl\{-3\,t\,
  \widetilde{{\Gamma}}/a^2\bigr\}
\end{equation}
and the $3\times 3$-matrix
\begin{equation}
  \boldsymbol{\mathsf{T}}(t,t'):=  \left(
    \begin{array}{ccc}
      1 & \int_{t'}^{t}{d}s\:\dot{\gamma}(s) & 0\\
      &1&0\\
      0&0&1\\
    \end{array}
  \right)\,. 
\end{equation}
Finally, the solution of the Zimm equation (\ref{zimm2}) is obtained by
inserting (\ref{solution}) in 
(\ref{koord}).

%
%
\section{Observables}\label{Sec3}
%
%

\subsection{Shear Stress}\label{Sec3.1}

We shall focus on the viscosity $\eta$ and the first and second normal
stress coefficients $\Psi^{(1)}$ and $\Psi^{(2)}$, respectively.
Therefore we need to compute the intrinsic shear stress
$\boldsymbol{\sigma}(t)$ as a function of the shear rate
$\dot{\gamma}(t)$.  Following Chap.~3 in Ref.\ \onlinecite{DoEd88} or
Chap.\ 16.3 in Ref.\ \onlinecite{BiCu87}, we express the shear stress
in terms of the force per unit area exerted by the polymers
\begin{equation}
  \label{kirkwood}
  \boldsymbol{\sigma}(t) = \lim_{t_{0}\to-\infty}
  -\frac{\rho_{0}}{N}\sum_{i=1}^{N} 
  \overline{\mathbf{F}_{i}(t) \mathbf{R}^{\dagger}_{i}(t)}.
\end{equation}
Here, $\mathbf{R}_{i}(t)$ is the solution of the equation of motion
(\ref{zimm2}) with some initial condition $\mathbf{R}_{i}(t_{0})$ at time
$t_{0}$ in the distant past (so that all transient effects stemming from the
initial condition have died out).  Moreover, $\rho_{0}$ stands for the monomer
concentration and $\mathbf{F}_{i}(t):=-\partial V/\partial \mathbf{R}_i(t)$ is
the net spring force acting on monomer $i$ at time $t$.  Using the
transformation (\ref{koord}) and the solution (\ref{solution}), it is readily
shown \cite{BrLoMu01a, BrLoMu01b} that the stress tensor (\ref{kirkwood}) is
given by
\begin{align}
  \label{linearresponse}
  \fl 
  \boldsymbol{\sigma}(t) =\chi(0) \, \boldsymbol{\mathsf{1}} +
  \int_{-\infty}^t {d} t' & \; \chi(t-t')\,
  \dot{\gamma}(t') \nonumber\\
  & \times \left(
    \begin{array}{ccc}
      2\,\int_{t'}^{t}{d}s\:\dot{\gamma}(s) & 1 & 0\\
      1&0&0\\
      0&0&0\\
    \end{array}
  \right)
\end{align}
for arbitrary strengths of the shear rate $\dot{\gamma}(t)$. Here, we
have defined the stress-relaxation function 
\begin{equation}
  \label{stressrelaxation}
  \chi(t) := 
  \frac{\rho_{0}}{N}\; {\rm Tr}
  \biggl[(\mathsf{1}-\widetilde{\mathsf{E}}_0)
  \exp\biggl(-\,\frac{6t}{a^{2}}\;  
  \widetilde{{\Gamma}}\biggr)\biggr]
\end{equation}
as a trace over the subspace of non-zero eigenvalues of
$\widetilde{{\Gamma}}$.

For a time-independent shear rate $\dot{\gamma}$, the shear stress
(\ref{linearresponse}) is also independent of time. The (intrinsic
zero-shear) viscosity $\eta$ is then related to shear stress via 
\begin{equation}
  \label{viscositydef}
  \eta :=\frac{\sigma_{x,y}}{\dot{\gamma}\rho_{0}}
\end{equation}
and the normal stress coefficients are defined by
\begin{equation}
  \label{normstressdef}
  \Psi^{(1)}:=\frac{\sigma_{x,x}-\sigma_{y,y}}{\dot{\gamma}^2\rho_0}\,,\qquad
  \Psi^{(2)}:=\frac{\sigma_{y,y}-\sigma_{z,z}}{\dot{\gamma}^2\rho_0}\,,
\end{equation}
respectively. Hence, the viscosity (\ref{viscositydef}) 
is given by
\begin{equation}
  \label{viscosity}
  \eta({\cal G}) = \frac{1}{\rho_0}\int_0^{\infty}{d}t\:\chi(t)=
  \frac{a^2}{3}\,
  \frac{1}{2N}
  \:{\rm Tr}\!\left[\frac{\mathsf{1}-\widetilde{\mathsf{E}}_0}
    {\widetilde{{\Gamma}}({\cal G})}\right] 
\end{equation}
for a fixed realization ${\cal G}$ of crosslinks. It is determined by the
trace of the Moore--Penrose inverse of $\widetilde{{\Gamma}}({\cal G})$.
According to (\ref{linearresponse}) and (\ref{normstressdef}), the second
normal stress coefficient $\Psi^{(2)}$ vanishes always, whereas
\pagebreak[1]
\begin{equation}
  \label{normalstress}
  \fl
  \Psi^{(1)}({\cal G})=\frac{2}{\rho_0}\int_0^{\infty}{d}t\,t\:\chi(t)=
  \left(\frac{a^{2}}{3}\right)^2
  \frac{1}{2N}
  \:{\rm Tr}\!\left[\frac{\mathsf{1}-\widetilde{\mathsf{E}}_0}
    {\bigl(\widetilde{{\Gamma}}({\cal G})\bigr)^{2}}\right].
\end{equation}
Again, we have made explicit the dependence on $\cal G$ in
(\ref{normalstress}). This will be convenient for computing the
average over all crosslink realizations in the next subsection.

\subsection{Disorder Average and Critical Behaviour}\label{Sec3.2}

Each crosslink realization ${\cal G}$ defines a random labelled graph on the
set of monomers, which can be decomposed into maximal path-wise connected
components or clusters
\begin{equation}
  \label{Eq2.22}
  {\cal G}=\bigcup_{k=1}^{K}{\cal N}_{k}\,.
\end{equation}
Here, ${\cal N}_{k}$ denotes the $k$-th cluster with $N_k$ monomers out of a
total of $K$ clusters (all depending on $\mathcal{G}$).  We also refer to
$N_{k}$ as the size of the cluster $\mathcal{N}_{k}$.  The associated modified
connectivity matrix $\widetilde{{\Gamma}}$ from (\ref{gammatilde}) is of
block-diagonal form with respect to the clusters.  Therefore one can decompose
any observable of the type $A(\mathcal{G}) = N^{-1} \mathrm{Tr}
f(\widetilde{\Gamma}(\mathcal{G}))$, where $f$ is some function on the reals,
into contributions from different clusters according to
\begin{equation}\label{Eq2.24}
  A({\cal G}) = \sum_{k=1}^{K} \frac{N_{k}}{N}\:A({\cal N}_{k})\,.
\end{equation}
Here, we have defined $A({\cal N}_{k}) := N_{k}^{-1} \mathrm{Tr}
f(\widetilde{\Gamma} (\mathcal{N}_{k})) $. In particular, (\ref{Eq2.24}) holds
for the viscosity (\ref{viscosity}) and for the first normal stress coefficient
(\ref{normalstress}).

In order to compute the average $\langle A \rangle$ of the observable $A$ over
all crosslink realizations in the macroscopic limit $M \to \infty$, $N \to
\infty$ with fixed crosslink concentration $c:=M/N$, we have to specify the
statistical ensemble that determines the realizations of crosslinks.  Two
distributions of crosslinks will be considered. (i)~Each pair of monomers is
chosen independently with equal probability $c/N$, corresponding to
Erd\H{o}s--R\'enyi random graphs, which are known to resemble the critical
properties of mean-field percolation. \cite{Ste77} After performing the
macroscopic limit, there is no macroscopic cluster for
$c<c_{\mathrm{crit}}=1/2$ and almost all clusters are trees. \cite{ErRe60}
Furthermore, all $n^{n-2}$ trees of a given size $n$ are equally likely.
(ii)~Clusters are generated according to three-dimensional continuum
percolation, which is closely related to the intuitive picture of gelation,
where monomers are more likely to be crosslinked when they are close to each
other. Since continuum percolation and lattice percolation are believed to be
in the same universality class, \cite{StAh94} we employ the scaling
description of the latter. It predicts \cite{StAh94} a cluster-size
distribution of the form
\begin{equation}
  \label{clustersizedist}
  \tau_n :=\biggl\langle{N}^{-1} \sum_{k=1}^{K} \delta_{N_{k},n}
\biggr\rangle \sim n^{-\tau} \exp\{-n/n^{*}\}
\end{equation}
for $\varepsilon:=(c_{\mathrm{crit}}-c) \ll 1$ and $n\to\infty$ with a
typical cluster size $n^{*}(\varepsilon) \sim \varepsilon^{-1/\sigma}$
that diverges as $\varepsilon \to 0$. Here, $\sigma$ and $\tau$ are
(static) critical exponents.

For the computation of the average $\langle A\rangle$ over all
crosslink realizations $\mathcal{G}$ it is convenient to introduce 
partial averages 
\begin{equation}
 \langle A \rangle_{n} := \tau_{n}^{-1}\biggl\langle
N^{-1}\sum_{k=1}^{K} \delta_{N_{k},n} \, A(\mathcal{N}_{k}) \biggr\rangle 
\end{equation}
of $A$ over all clusters of a given size $n$.  Using (\ref{Eq2.24}) and
reordering the clusters, one gets the identity
\begin{equation}
  \label{reordering}
  \langle A \rangle 
  = \biggl\langle
  \sum_{k=1}^{K} \frac{N_{k}}{N}\; A(\mathcal{N}_{k}) \biggr\rangle  
  = \sum_{n=1}^{\infty} n\tau_{n}\langle A \rangle_{n} \,,
\end{equation}
which is valid in the absence of an infinite cluster.  Now suppose one has the
scaling
\begin{equation}
  \label{partialav}
  A_{n} :=  \langle A\rangle_{n}\big|_{\varepsilon =0}\sim n^b
\end{equation}
of the partial average at the critical point as $n\to\infty$.
Due to the absence of relevant scales at the critical point, this is quite a
natural behaviour. Then, (\ref{reordering}) and (\ref{clustersizedist}) imply
the critical divergence 
\begin{equation}
  \label{critscaling}
  \fl
  \langle A\rangle \sim\varepsilon^{-u}\quad\text{as}\quad\ep\downarrow
  0,\quad\text{with}\quad u=(2-\tau+b)/\sigma 
\end{equation}
for the crosslink-averaged observable $A$, provided that $u>0$. 
We will therefore study the scaling
\begin{equation}
  \label{ennscaling}
  \eta_{n}\sim n^{b_{\eta}}\quad\text{and}\quad
  \Psi^{(1)}_{n}\sim n^{b_{\Psi}}
\end{equation}
as $n\to\infty$ to explore critical rheological behaviour at the
gelation transition within the Zimm model. 

Formulas like (\ref{reordering}) -- (\ref{ennscaling}) may also be familiar
from scaling theories for gelation. We go beyond such approaches in that we
have mapped the dynamical properties of a gelling molecular system to
a percolation problem, see e.g.\ (\ref{viscosity}) and
(\ref{normalstress}). This mapping has been fully
derived within a (semi-) microscopic dynamical model, the Zimm model
for randomly crosslinked monomers, and not merely postulated from
\emph{adhoc} assumptions, as is usually done in scaling theories.
In the following section we describe the numerical solution of the
percolation problem.

%
%
\section{Numerical Results}\label{Sec4}
%
%
\subsection{Erd\H{o}s--R\'enyi Random Graphs}\label{Sec4.1}

For numerical purposes it is convenient to compute the eigenvalues of
the non-symmetric matrix
${\widehat{\Gamma}}:=\mathsf{H}^{\mathrm{eq}}{\Gamma}$ rather than
those of ${\widetilde{\Gamma}}=(\mathsf{H}^{\mathrm{eq}})^{1/2}
{\Gamma}(\mathsf{H}^{\mathrm{eq}})^{1/2}$ because this prevents us from
computing the expensive square root of $\mathsf{H}^{\mathrm{eq}}$. The
fact that ${\widetilde{\Gamma}}$ and ${\widehat{\Gamma}}$ have the
same eigenvalues can easily be proven by observing that if
$\psi$ is an eigenvector of ${\widetilde{\Gamma}}$ with corresponding
eigenvalue $\lambda$ then $(\mathsf{H}^{\mathrm{eq}})^{\pm 1/2}\,\psi$
is a right/left eigenvector of ${\widehat{\Gamma}}$ with the same
eigenvalue $\lambda$.

\begin{figure*}
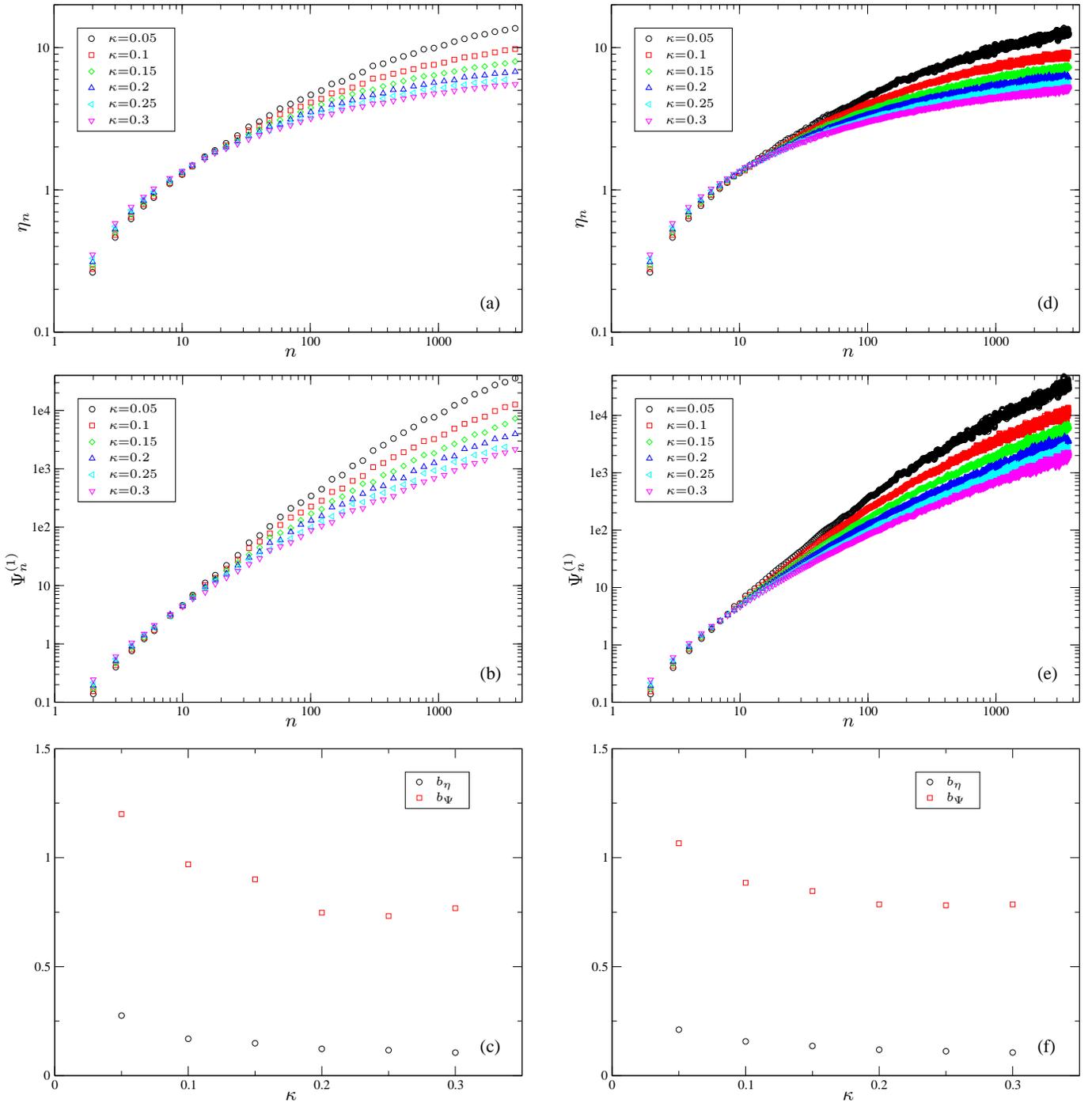

  \begin{center}
    {\leavevmode 
      \psfrag{Tr(1-E/Gamma)/N}{${\eta}_n$} 
      \psfrag{n}{${n}$}
      \psfrag{kappa=0-05}{$\scriptstyle\kappa =0.05$}
      \psfrag{kappa=0-1}{$\scriptstyle\kappa =0.1$}
      \psfrag{kappa=0-15}{$\scriptstyle\kappa =0.15$}
      \psfrag{kappa=0-2}{$\scriptstyle\kappa =0.2$}
      \psfrag{kappa=0-25}{$\scriptstyle\kappa =0.25$}
      \psfrag{kappa=0-3}{$\scriptstyle\kappa =0.3$}
      \epsfig{file=\picdirectory/viscMF.eps, clip=,
        width=.99\columnwidth}    
    \hfill
      \psfrag{Tr(1-E/Gamma)/N}{${\eta}_n$}
      \psfrag{n}{${n}$}
      \psfrag{kappa=0-05}{$\scriptstyle\kappa =0.05$}
      \psfrag{kappa=0-1}{$\scriptstyle\kappa =0.1$}
      \psfrag{kappa=0-15}{$\scriptstyle\kappa =0.15$}
      \psfrag{kappa=0-2}{$\scriptstyle\kappa =0.2$}
      \psfrag{kappa=0-25}{$\scriptstyle\kappa =0.25$}
      \psfrag{kappa=0-3}{$\scriptstyle\kappa =0.3$}
      \epsfig{file=\picdirectory/smoothedViscPerc.eps, clip=,
        width=.99\columnwidth} 
      \par\medskip
      \psfrag{Tr(1-E/Gamma)/N}{$\Psi^{(1)}_n$} \psfrag{n}{${n}$}
      \psfrag{kappa=0-05}{$\scriptstyle\kappa =0.05$}
      \psfrag{kappa=0-1}{$\scriptstyle\kappa =0.1$}
      \psfrag{kappa=0-15}{$\scriptstyle\kappa =0.15$}
      \psfrag{kappa=0-2}{$\scriptstyle\kappa =0.2$}
      \psfrag{kappa=0-25}{$\scriptstyle\kappa =0.25$}
      \psfrag{kappa=0-3}{$\scriptstyle\kappa =0.3$}
      \epsfig{file=\picdirectory/normStressMF.eps, clip=,
        width=.99\columnwidth}
      \hfill
      \psfrag{Tr(1-E/Gamma)/N}{$\Psi^{(1)}_n$} \psfrag{n}{${n}$}
      \psfrag{kappa=0-05}{$\scriptstyle\kappa =0.05$}
      \psfrag{kappa=0-1}{$\scriptstyle\kappa =0.1$}
      \psfrag{kappa=0-15}{$\scriptstyle\kappa =0.15$}
      \psfrag{kappa=0-2}{$\scriptstyle\kappa =0.2$}
      \psfrag{kappa=0-25}{$\scriptstyle\kappa =0.25$}
      \psfrag{kappa=0-3}{$\scriptstyle\kappa =0.3$}
      \epsfig{file=\picdirectory/smoothedNormPerc.eps, clip=,
        width=.99\columnwidth}
      \par\medskip
      \psfrag{n}{$\kappa$}
      \psfrag{fitExponent}{} \psfrag{b_eta}{$\scriptstyle b_{\eta}$}
      \psfrag{b_psi}{$\scriptstyle b_{\Psi}$}
      \epsfig{file=\picdirectory/fit_viscNormStressMF.eps, clip=,
        width=.99\columnwidth}
      \hfill
      \psfrag{n}{$\kappa$}
      \psfrag{fitExponent}{} \psfrag{b_eta}{$\scriptstyle b_{\eta}$}
      \psfrag{b_psi}{$\scriptstyle b_{\Psi}$}
      \epsfig{file=\picdirectory/fit_viscNormPerc.eps, clip=,
        width=.99\columnwidth} 
    \caption{Numerical data to determine the scaling
      (\ref{ennscaling}) for random clusters in the case of
      Erd\H{o}s--R\'enyi random graphs (left column) and
      three-dimensional bond percolation (right column). 
      In each case the averaged viscosity
      ${\eta}_n$ (top) and normal stress coefficient $\Psi^{(1)}_n$
      (middle) are plotted for different strengths of the hydrodynamic
      interaction parameter $\kappa$ as a function of the cluster size $n$
      on a double logarithmic scale. Power-law fits to the data
      yield the exponents $b_{\eta}$ and $b_{\Psi}$ as
      a function of $\kappa$ (bottom).
      \label{fig:1}
    }
    }
  \end{center}
\end{figure*}

As already mentioned in Sec.~\ref{Sec3.2}, the average 
$\langle\bullet\rangle_{n}$ extends over all $n^{n-2}$ equally weighted
labelled trees of size $n$ in the case of Erd\H{o}s--R\'enyi
random graphs and, hence, is independent of the crosslink
concentration $c$. Random labelled
trees of a given size have been generated via the Pr\"ufer algorithm
and handled with the {\sc LEDA} library. \cite{MeNa99} The preaveraged
mobility matrix (\ref{preav}) is computed with the function $h$
from (\ref{preavrotne}), corresponding to the Rotne--Prager--Yamakawa
tensor. The resistances 
${\cal R}_{i,j}$ in trees reduce to shortest paths, that is graph distances,
which are calculated with the Dijkstra algorithm. \cite{MeNa99} The
eigenvalues of $\widehat{\Gamma}$ are then computed with the {\sc
  LAPACK} library. For suitable, logarithmically equidistant cluster sizes
$n\in[2,4000]$ we average the viscosity and the normal stress coefficient 
over $50$ trees, which turned out to yield an acceptable
computer-time/accuracy trade-off. In Figs.~\ref{fig:1}(a) and~(b) we plot
${\eta}_{n}$ and $\Psi^{(1)}_n$ as a function of $n$ on a
double-logarithmic scale for different values of the hydrodynamic
interaction parameter $\kappa$. According to (\ref{ennscaling}) the
exponents $b_{\eta}$ and $b_{\Psi}$ 
are obtained by power law fits in the
large $n$-range, for which we choose the interval
$n\in[700,4000]$, see Fig.~\ref{fig:1}(c). For the viscosity the exponent
decreases from $b_{\eta}=0.28$ for $\kappa=0.05$ to $b_{\eta}=0.11$ for
$\kappa=0.3$. The Rouse exponent for $\kappa=0$ is exactly given by
\cite{BrLoMu01a} $b_{\eta}=1/2$. 
The exponent $b_{\Psi}$ of the normal stress coefficient ranges from
$b_{\Psi}=1.2$ for $\kappa=0.05$ to $b_{\Psi}=0.73$ for $\kappa=0.25$.
The Rouse value for $\kappa=0$ is exactly given \cite{BrLoMu01b} by
$b_{\Psi}=2$. 

\subsection{Three-Dimensional Percolation}\label{Sec4.2}

For the generation of clusters according to three-dimensional {\it bond
  percolation} we apply the Leath Algorithm. \cite{Lea76} It generates a
sequence $\{{\cal N}_l\}_{l=1}^L$ of clusters, in terms of which the disorder
average is readily computed via $\langle A\rangle=\lim_{L\to\infty} L^{-1}
\sum_{l=1}^LA({\cal N}_l)$.  This implies $\langle A\rangle_n =
\lim_{L\to\infty} \sum_{l=1}^L\delta_{N_l,n} A({\cal
  N}_l)/\sum_{l=1}^L\delta_{N_l,n}$ for the average over clusters of size $n$.
The algorithm has been tested by verifying the scaling of the cluster-size
distribution $\tau_n$. Second, for small values of $n$, we compared the number
of clusters with known exact values.\cite{SyGaGl81} Third, we verified that
the exponent $2/d_f$, which governs the scaling of the squared radius of
gyration as a function of cluster size $n$ at the critical point,\cite{StAh94}
comes out as $2/2.53$ from the simulation. For each generated cluster the
resistances ${\cal R}_{i,j}$ are computed from the Moore--Penrose inverse
$\mathsf{Z}$ of the connectivity matrix $\mathit{\Gamma}$ -- see below
Eq.~(\ref{preav}) -- and inserted into (\ref{preav}) with $h$ corresponding to
the Rotne--Prager--Yamakawa tensor.  The eigenvalues of $\widehat{\Gamma}$ are
then computed with the {\sc LAPACK} library.  We were forced to restrict
cluster sizes to values $n<4000$ due to the limited amount of memory, which is
required for the generation and diagonalization of the matrix product
$\widehat{\Gamma}=\mathsf{H}^{\mathrm{eq}}{\Gamma}$. Moreover, for calculating
disorder averages we restrict the number of realizations pertaining to a given
cluster size to a maximum of $50$. However, within the present numerical
effort this maximum number is not even attained for larger cluster sizes.
Therefore the disorder averaged quantities are still subject to fluctuations.
In order to obtain smooth curves for $\eta_{n}$ and $\Psi_{n}$ we have also
smoothed out the raw data by performing a running average over cluster sizes
in the window $[n-5,n+5]$. The thus obtained values for $\eta_{n}$ and
$\Psi_n$ are plotted in Figs.~\ref{fig:1}(d) and (e), respectively, as a
function of $n$ on a double-logarithmic scale for different values of
$\kappa$. The exponents $b_{\eta}$ and $b_{\Psi}$, extracted by fitting the
curves in Figs.~\ref{fig:1}(d) and (e) to a power law in the interval
$n\in[800,4000]$, are shown in Fig.~\ref{fig:1}(f). The numerical values for
$b_{\eta}$ are nearly identical to those obtained for Erd\H{o}s--R\'enyi
random graphs.  Again, one observes a decrease from $b_{\eta}=0.21$ for
$\kappa=0.05$ to $b_{\eta}=0.11$ for $\kappa=0.3$. The exponent $b_{\Psi}$ of
the normal stress coefficient ranges from $b_{\Psi}=1.1$ for $\kappa=0.05$ to
$b_{\Psi}=0.78$ for $\kappa=0.25$. The corresponding Rouse values for
$\kappa=0$ follow from exact analytical arguments
\cite{BrLoMu99,BrLoMu01a,Mul03} and are given by $b_{\eta} = (2/d_{s}) -1
\approx 1/2$ and $b_{\Psi} = (4/d_{s}) -1 \approx 2$, respectively. Here,
$d_{s} \approx 4/3$ is the spectral dimension of the incipient percolating
cluster, whose numerical value is very well approximated by the
Alexander--Orbach conjecture.\cite{StAh94}

\subsection{Ring Polymers}\label{Sec4.3}

We suspect that the observed variation of the exponent values with $\kappa$
may be due to crossover and finite-size effects. To clarify this question it
is useful to study a system where the exponents are known analytically.
Therefore we (re-)investigate the viscosity $\eta_{\rm{ring}}$ and the first
normal stress coefficient $\Psi^{(1)}_{\rm{ring}}$ of ring polymers in the
Zimm model with the Rotne--Prager--Yamakawa tensor. The scaling of both
quantities with ring size $n$ as $n\to\infty$ can be deduced from
long-standing analytical results, \cite{BlZi66} which lead to
$b_{\eta,\mathrm{ring}} = 1/2$ and $b_{\Psi,\mathrm{ring}} =2$. We focus here
on the onset of this asymptotic behaviour and how it is affected by crossovers
for different $\kappa$. This provides us with a reference system when
discussing the scaling of ${\eta}_n$ and $\Psi^{(1)}_n$ in the case of random
clusters in Section~\ref{Sec5}.

\begin{figure}[H]
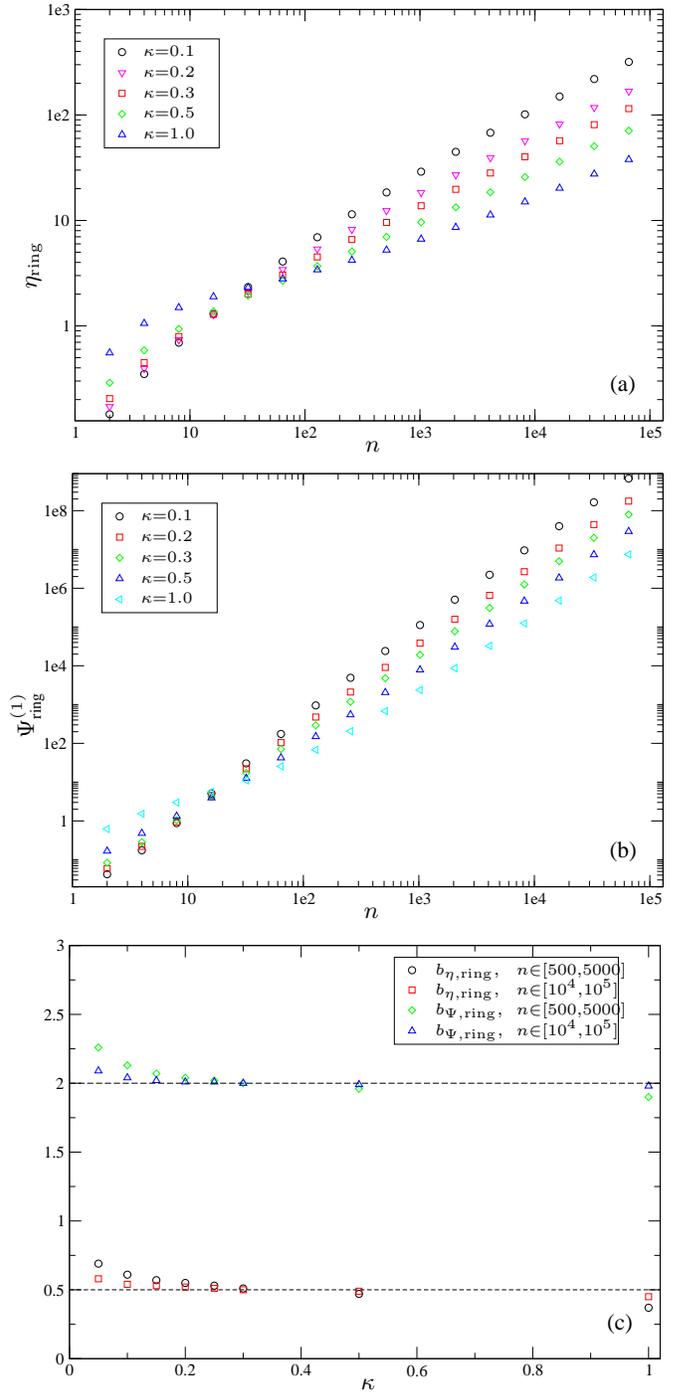

  \begin{center}
    {
  {
    \psfrag{Tr(1-E/Gamma)/N}{$\eta_{\rm{ring}}$} \psfrag{n}{${n}$}
    \psfrag{kappa=0-1}{$\scriptstyle\kappa =0.1$}
    \psfrag{kappa=0-2}{$\scriptstyle\kappa =0.2$}
    \psfrag{kappa=0-3}{$\scriptstyle\kappa =0.3$}
    \psfrag{kappa=0-5}{$\scriptstyle\kappa =0.5$}
    \psfrag{kappa=1-0}{$\scriptstyle\kappa =1.0$}
    \epsfig{file=\picdirectory/ringVisc.eps, clip=,
      width=.99\columnwidth}
    \par\medskip
  }
  { \psfrag{n}{$n$}
    \psfrag{Tr(1-E_0/G^2)/N}{$\Psi^{(1)}_{\text{ring}}$}
    \psfrag{kappa=0-1}{$\scriptstyle\kappa =0.1$}
    \psfrag{kappa=0-2}{$\scriptstyle\kappa =0.2$}
    \psfrag{kappa=0-3}{$\scriptstyle\kappa =0.3$}
    \psfrag{kappa=0-5}{$\scriptstyle\kappa =0.5$}
    \psfrag{kappa=1-0}{$\scriptstyle\kappa =1.0$}
    \epsfig{file=\picdirectory/ringNormStress.eps, clip=,
      width=.99\columnwidth}
    \par\medskip
  }
  { 
    \psfrag{n}{$\kappa$} 
    \psfrag{b_eta in [500,5000] blanks}{$\scriptstyle 
      b_{\eta,\mathrm{ring}},\;\;\; n \in [500,5000]$} 
    \psfrag{b_psi in [500,5000]}{$\scriptstyle
      b_{\Psi,\mathrm{ring}},\;\:\, n \in [500,5000]$} 
    \psfrag{b_eta in [10^4,10^5]}{$\scriptstyle b_{\eta,\mathrm{ring}},\;\;\; 
      n \in [10^{4},10^{5}]$} 
    \psfrag{b_psi in [10^4,10^5]}{$\scriptstyle
      b_{\Psi,\mathrm{ring}},\;\:\, n \in [10^{4},10^{5}]$}
    \psfrag{fitExponent}{}
    \epsfig{file=\picdirectory/fit_ringViscNormStress.eps, clip=,
      width=.99\columnwidth} 
  }
      \caption{Numerical data to determine the scaling (\ref{ennscaling}) for
        ring polymers. The viscosity ${\eta}_{\mathrm
          {ring}}$ (a) and the normal stress coefficient
        ${\Psi}^{(1)}_{\mathrm {ring}}$ (b) are plotted for different
        strengths of the hydrodynamic interaction parameter $\kappa$
        as a function of the cluster size $n$ on a double logarithmic
        scale. (c) shows the exponents $b_{\eta,\mathrm{ring}}$ and 
        $b_{\Psi,\mathrm{ring}}$ from power-law fits to the data of
        (a) and (b). The fits were performed for two different ranges
        of cluster sizes $n$. Additional data points for
        $\kappa=0.05, 0.15$ and $ 0.25$ in (c) stem from curves which have
        been omitted in (a) and (b) for reasons of clarity. The two horizontal
        lines indicate the exact values $b_{\eta,\mathrm{ring}} = 1/2$ and
        $b_{\Psi,\mathrm{ring}} =2$.
        \label{fig:2}}
    }
  \end{center}
\end{figure}

Due to the cyclic structure of a ring polymer the associated matrices
$\mathsf{H}^{\rm{eq}}$ and $\Gamma$ are circulant matrices.  Hence, they are
simultaneously diagonalizable. In fact, the $j$-th component of the $l$-th
eigenvector of $\widehat{\Gamma}$ for a ring of size $n$ is explicitely given
by $\psi^{(l)}_j=\exp(i2\pi j l/n)$, and as a result the eigenvalues can be
written in terms of Fourier transforms.  \cite{OeZy92} Therefore,
$\eta_{\rm{ring}}$ and $\Psi^{(1)}_{\rm{ring}}$ are efficiently computed by
Fast Fourier Transformation up to ring sizes $n=10^5$.  The resulting
viscosity and the first normal stress coefficient are shown in
Figs.~\ref{fig:2}(a) and (b) on a double logarithmic scale. The data is then
fitted to a power law in two different fit ranges. In addition to a fit in the
terminal large-$n$ range, $n\in[10^4,10^5]$, we performed a second fit in the
range $n\in[500,5000]$, which is roughly where we had to do the fits in the
random-cluster case. The fit exponents are shown in Fig.~\ref{fig:2}(c).
Apparently, they depend on the fit range. For $\kappa=0.05$ we find
$b_{\eta,\mathrm{ring}}=0.69$ from the small-$n$ fit. This value clearly
exceeds the theoretical one $b_{\eta,\mathrm{ring}}=1/2$. Even the
corresponding value $b_{\eta,\mathrm{ring}}=0.58$ from the large-$n$ fit still
has an error of $36\%$. In contrast, for $\kappa=0.3$ both values,
$b_{\eta,\mathrm{ring}}=0.51$ and $0.50$, are quite close to the exact one.

In fact, given the Fourier representation of the eigenvalues of
$\widehat{\Gamma}$, it is straightforward to demonstrate the
occurrence of a crossover at $n \approx \pi/\kappa^{2}$ from Rouse behaviour,
$\eta_{\mathrm{ring}} \sim n$, to the asymptotic Zimm behaviour
$\eta_{\mathrm{ring}} \sim n^{1/2}/\kappa$ for all $n \gg \kappa^{-2}$. Hence,
the larger $\kappa$, the less important is residual Rouse behaviour in the
numerical data for the scaling of $\eta_{\mathrm{ring}}$. The same holds true
for $\Psi^{(1)}_{\mathrm{ring}}$. Unfortunately, choosing larger values for
$\kappa$ is not a practicable way out for obtaining good-quality data. This is
because for large $\kappa$
the asymptotics 
\begin{equation}
  \label{nonoseen}
  h(x) \sim 1 - (\pi x)^{-1/2}
\end{equation}
of (\ref{preavrotne}) as $x \to \infty$ becomes noticible and leads to the
transient behaviour $\eta_{\mathrm{ring}} \sim \kappa n^{0}$ for intermediate
$n$. We have observed such a behaviour for (unphysically large) $\kappa >10$
(not shown).  But even the data for $\kappa =0.5$ and $\kappa =1.0$ in
Fig.~\ref{fig:2} are still slightly influenced by (\ref{nonoseen}).

In summary, whereas there is a generic overestimate of the scaling exponents
for small $\kappa$ due to residual Rouse behaviour, the exponents are
underestimated for higher $\kappa$ due to the asymptotics (\ref{nonoseen}).
The optimal value for minimal finite-size effects in ring polymers appears to
be $\kappa \approx 0.3$ in Fig.~\ref{fig:2}(c).

%
%
\section{Discussion}\label{Sec5}
%
%

Using the Zimm model for randomly crosslinked monomers, we have
determined the scaling (\ref{ennscaling}) of the averaged viscosity
and of the averaged first normal stress coefficient over clusters of a
given size $n$. Figs.~\ref{fig:1}(c) and~(f) display a crossover from
the Rouse values at $\kappa=0$ to the Zimm values at non-zero
$\kappa$. We estimate the latter as
\begin{equation}
  \label{zimmexp}
  b_{\eta} \approx 0.11 \qquad\mathrm{and}\qquad b_{\Psi} \approx 0.77\,,
\end{equation} 
from our data for $\kappa=0.3$. A detailed discussion of this choice of
$\kappa$ and of possible origins of the dependence of $b_{\eta}$ and
$b_{\Psi}$ on $\kappa$ will be given below. Within the accuracy of our data,
the exponents are the same for both Erd\H{o}s--R\'enyi random graphs and
three-dimensional percolation.

The critical behaviour of the averaged viscosity $\langle\eta\rangle \sim
\varepsilon^{-k}$ and of the averaged first normal stress coefficient
$\langle\Psi^{(1)} \rangle\sim \varepsilon^{-\ell}$ for a polydisperse
gelling solution of crosslinked monomers then follows from (\ref{zimmexp})
and (\ref{critscaling}). For the viscosity this implies a {\it finite} value
at the gel point for both, Erd\H{o}s--R\'enyi random graphs and
three-dimensional bond percolation. In contrast, the first normal
stress coefficient is found to diverge with an exponent that depends
on the cluster statistics. Choosing the cluster statistics according
to Erd\H{o}s--R\'enyi random graphs, we find $\ell \approx 0.54$.
The case of three-dimensional bond percolation leads to the higher
value $\ell \approx 1.3$. These exponent values are less than a third
in magnitude than the corresponding exact analytical predictions $\ell
=3$, respectively $\ell \approx 4.1$ of the Rouse model for randomly
crosslinked monomers \cite{BrLoMu01b, BrMuZi02, Mul03} with the
corresponding cluster statistics.

The dependence of the critical exponents $b_{\eta}$ and $b_{\Psi}$ on the
hydrodynamic interaction strength $\kappa$ in Figs.~\ref{fig:1}(c) and~(f) may
be due to finite-size effects.  In particular the onset of the true asymptotic
regime of these quantities may depend on $\kappa$.
In order to better understand finite-size effects, we have examined the Zimm
dynamics of polymer rings in Sec.~\ref{Sec4.3} and determined the scaling of
the viscosity $\eta_{\mathrm{ring}} \sim n^{b_{\eta,\mathrm{ring}}}$ and of
the first normal stress coefficient $\Psi^{(1)}_{\mathrm{ring}} \sim
n^{b_{\Psi, \mathrm{ring}}}$ with the ring size $n$. For rings one can access
much higher values of $n$ as for random clusters, see Figs.~\ref{fig:2}(a)
and~(b). In particular, the exactly known scaling exponents
$b_{\eta,\mathrm{ring}} =1/2$ and $b_{\Psi, \mathrm{ring}} =2$, which are
universal in $\kappa >0$, can be extracted from our data in
Fig.~\ref{fig:2}(c). However, if we did not exploit the full range of
available ring sizes and restricted the fit to those lower values of $n$ which
could also be accessed for random clusters, then universality would be veiled
by finite-size effects. Finite-size effects are more pronounced for $\kappa
\le 0.15$ and $\kappa > 0.5$. Thus, we conclude (i)~that the random-cluster
data have not reached either the asymptotic large-$n$ regime yet for $\kappa
\le 0.15$ in Fig.~\ref{fig:1}, (ii)~that the asymptotic regime \emph{is}
universal and (iii)~that the data for $\kappa = 0.3$ should be the most
reliable ones.

The exponent $b_{\eta}$ has also been investigated experimentally.  In
Ref.~\onlinecite{MaOhOn86} measurements on randomly branched
polystyrenes have been performed, resulting in $b_{\eta} \in
[0.2,0.25]$. Measurements on branched polyethyleneimine \cite{PaCh96}
yield the slightly higher value $b_{\eta} \approx 0.31$.
Brownian-dynamics simulations of hyperbranched polymers were performed
in Ref.~\onlinecite{ShAdLy02}. They also account for
\emph{fluctuating} hydrodynamic interactions corresponding to $\kappa
= 0.35$, as well as for excluded-volume interactions and lead to
$b_{\eta} = 0.13$. This result is remarkably close to our finding
$b_{\eta} \approx 0.11$ for the highest coupling strength $\kappa
=0.3$ that we have considered, whereas the experimental findings are
consistently above our value (see the discussion below).

Next we compare our findings with the scaling argument which is summarized in
Eq.~(\ref{zimmscaling}). For phantom clusters, {\it i.e.}\ in the absence of
excluded-volume interactions, the Hausdorff fractal dimension is equal to the
Gaussian fractal dimension \cite{Cat85,Vil88} $d^{({\mathrm G})}_{f} :=
2d_s/(2-d_s)$, where $d_s$ is the spectral dimension. Here we estimate
$d_{s}\approx 4/3$ according to the Alexander--Orbach conjecture, which is
known to be an excellent approximation albeit not being exact. For $d=3$
Eq.~(\ref{zimmscaling}) then
implies $b_{\eta} \approx -1/4$, and for $d=6$ one has $b_{\eta} \approx 1/2$.
The latter value corresponds to Erd\H{o}s--R\'enyi random graphs, whose
critical properties are identical to those of mean-field percolation.
\cite{Ste77} Both values can be definitely ruled out by our data. Thus we
conclude that the scaling relation (\ref{zimmscaling}) does not apply to the
Zimm model for randomly crosslinked monomers. This failure comes as a surprise
because it is known from a recent investigation of diffusion constants within
this model \cite{KuLoMu03} that the exact results are in accordance with long
standing scaling relations when inserting $d^{({\mathrm G})}_{f}$ for $d_{f}$.

Coming back to the experimental $k$-values listed in Table~\ref{tab:1} and
considering also the exact prediction $k = (1-\tau +2/d_{s})/\sigma \approx
0.71$ of the Rouse model for gelling monomers, \cite{BrLoMu99, BrLoMu01a,
  Mul03} we conclude that an explanation for the broad scatter of the data in
the literature calls for additional relevant interactions than those accounted
for in the Zimm or Rouse model. This may be due to the preaveraging
approximation.  In particular, it throws away hydrodynamic interactions among
different clusters.  But we do not expect this to be the sole relevant
simplification of the Zimm model, because linear polymers show a decrease in
the viscosity when abandoning the preaveraging approximation,\cite{Fix81} and
effects of preaveraging for branched molecules are even more pronounced than
those for linear ones.  \cite{BuScSt80} Rather it seems that there are no
satisfactory explanations without considering excluded-volume interactions.
Indeed, simulations \cite{GaArCo00} of the bond-fluctuation model deliver
higher values $k\approx 1.3$ in accordance with the scaling relation $k=2\nu
-\beta$, which arises from heuristically merging Rouse-type and
excluded-volume properties.


\begin{acknowledgments}
  We acknowledge financial support by the DFG
  through SFB~602 and Grant No.\ Zi~209/6--1.
\end{acknowledgments}

\end{document}